# Classification of Pneumonia and Tuberculosis from Chest X-rays

1st Muhammad Abubakar
*Department of Computing and technology.*
*Abasyn University Islamabad Campus*
Islamabad, Pakistan
m.abubakar.asif@gmaill.com

2nd Inamullah Shah
*Department of Computing and technology.*
*Abasyn University Islamabad Campus*
Islamabad, Pakistan
Inamullahshah1489@gmail.com

3rd Waqar Ali
*Department of Computing and technology.*
*Abasyn University Islamabad Campus*
Islamabad, Pakistan
waqar.ali@abasynisb.edu.pk

4th Farrukh Bashir
*Department of Computing and technology.*
*Abasyn University Islamabad Campus*
Islamabad, Pakistan
farrukh.bashir@abasynisb.edu.pk

*Abstract*— Artificial intelligence (AI) and specifically machine learning is making inroads into number of fields. Machine learning is replacing and/or complementing humans in a certain type of domain to make systems perform tasks more efficiently and independently. Healthcare is a worthy domain to merge with AI and Machine learning to get things to work smoother and efficiently. The X-ray based detection and classification of diseases related to chest is much needed in this modern era due to the low number of quality radiologists. This thesis focuses on the classification of Pneumonia and Tuberculosis – two major chest diseases – from the chest X-rays. This system provides an opinion to the user whether one is having a disease or not, thereby helping doctors and medical staff to make a quick and informed decision about the presence of disease. As compared to previous work our model can detect two types of abnormality. Our model can detect whether X-ray is normal or having abnormality which can be pneumonia and tuberculosis 92.97% accurately.

*Keywords—Chest X-rays, Medical Imaging, Pneumonia, Tuberculosis, Machine Learning, Artificial Intelligence*

## I. INTRODUCTION

Medical Image Classification systems are winning the market and proved to be a game-changer in the medical treatments. Big 4 giants invested money in the health sector to improve the outcomes and results of health care systems. Siemens the Deutsch Company has a separate department of intelligent healthcare which are working on visualizations of the human internal body and classify diseases from it.

The discovery of X-rays in 1895 by Wilhelm Roentgen led to the first Nobel Prize in Physics. Computed Tomography ranks as one of the top five medical developments in the last 40 years, according to most medical surveys. It has proven as valuable as a medical diagnostic tool that the 1979 Nobel Prize in Medicine was awarded to the inventors of CT.

A classification system is trained enough to work in the future. Scientists and engineers are making it diverse to identify every type of scenario to give the best results. But in this case, it is always an issue for a patient to believe in a machine, so classification systems propose their results. But still, they are helping doctors and radiologists to save time. Nowadays Medical Imaging techniques are used to identify different diseases like Tumors from MRI Cardiomegaly from Echo also chest diseases form chest X-rays. Kaggle one of the biggest data science competition platform is providing scientists stacks of Medical Imaging data and tasks to train algorithms and everyone competes to give remarkable results. So medical imaging is now a game-changer. In developed countries mostly hospital hire fewer doctors and invest more on Intelligent smart health equipment which helps them work faster convenient and efficient they also produce better results.

Besides this all Medical Image classification systems help doctors to identify disease faster and save time to help them work on focused areas, determining which surgeries are necessary, also it improves the patient placement into appropriate areas of care, such as ICU.

In the following points given below we discussed the significance of our work.

Our work is associated with the detection of abnormalities in chest X-rays.

The model we created can detect pneumonia, tuberculosis, and decide our X-ray is normal having no abnormality. As compared to previous work mostly work is focused in pneumonia.

The architecture of our model is small in size and is performing relatively better than other algorithms.

## II. RELATED WORK

### A. Research Review

The author and his team used CNN Model for classifying tuberculosis in chest X Rays. The dataset used in this is obtained from Peruvian partners at "Scio's en Salud". The dataset contains 4701 images in which 453 are labeled as normal and 4258 labeled as abnormal. The final accuracy found after Alex net is about 85.68% a significant improvement from non-shuffle sampling which is 53.02%. Accuracy is achieved 85% only on TB [1].

Identify applicable funding agency here. If none, delete this text box.

Do-un Jeoun author of this article used dataset of size 112,120 the images are transformed from 1024x1024 to 224x224 for extracting features from the author of the images used Densenet-169 architecture and pass the output to Support vector machines (SVM) to predict variables achieved accuracy is 80%. They applied different techniques and highest accuracy achieved is 80% [2].

In this article, Wei dai used the JSRT dataset in this research and the proposed methodology is to identify the x-ray of the patient whether it is normal or not by using organ segmentation techniques and then identify the report using SCAN the main drawback of their work is they use a very small amount of data. Dataset size is too much small mostly used is synthetic data [3].

The researcher and his fellows used Chest X-ray 14 dataset, containing over 100,000 frontal view X-ray images, published by the National Institutes of Health. The proposed methodology by the researcher is that he used the Chex net algorithm; it is a state-of-the art machine-learning algorithm to detect pneumonia at a level of a human practicing radiologist. It is a 121-layer Convoluted Neural Classification of Pneumonia and Tuberculosis from Chest X-rays 10 Network h. The Chex Net algorithm can identify 14 pathologies from a Chest X-ray. The performance of Chex Net reported an F1 score of 0.435. In this article, Wei dai used the JSRT dataset in this research and the proposed methodology is to identify the x-ray of the patient whether it is normal or not by using organ segmentation techniques and then identify the report using SCAN the main drawback of their work is they use a very small amount of data. Testing score is quite low in this case [4].

Former Apple engineer David W.Dai worked on detection of Pneumonia and tuberculosis. He used JSRT and Montgomery Chest set. He used very minute dataset and generate synthetic data and trained model on that synthetic dataset. The issue is in their dataset it is small and mostly synthetic images are added in it [5][6].

### B. Existing Systems in production

There are existing several platforms which provide proper Diseases Classification using Medical Images according to the data on which it is trained. Review of some of the famous Systems are given below.

The Chex Net is an AI based model which is has capability to detect pneumonia from CXR which is equivalent to Radiologist level. The team trained and developed the model that can classify pneumonia from CXR at a level experienced radiologist. Chex Net can detect all 14 diseases from chest X-rays which can be only identified in chest x-rays and achieve state of the art results on all 14 diseases[7].

Xray4all is a web-based system for users to provide chest X-rays detailed results after diagnosing them. It is trained on Chex Pert dataset provided by Stanford University. It supports multiple tasks like X-rays and Histopathology slide analysis. This system analyzes uploaded images on a secure cloud backend and provides a probabilistic interpretation for different medical conditions[8].

### III. DATASET

Dataset collection and development is the first step towards training the ML algorithm. We need some data on which we need to train our Algorithms to perform classification tasks. In our case, we need X-ray Images to train our algorithms because we are detecting the diseases from the Xrays.

It is very difficult to collect Medical data because it is secure and needs lots of paperwork to be done. But there are many researchers and scientists published medical data for us to make use of that data by generating useful insights. So, we collected our data from different platforms related to Medics. The following are the platforms that provided us data

- Montgomery County X-ray Set
- Shenzhen China set
- Kaggle

### A. Dataset Count

We labelled our data into three classes which are shown in table 1. In Normal we have normal images of chest X-rays having no abnormality[9]. In pneumonia we added the images of pneumonia the images were taken from Kaggle dataset [10][10]. In tuberculosis we used the tb images from the Montgomery and Shenzhen china dataset[9].

*Table 1 Dataset Count*

| Normal | Pneumonia | Tuberculosis |
|--------|-----------|--------------|
| 1989   | 4273      | 394          |

### B. Pre-processing

There are some issues we face during the pre-processing step. The main issues we face in these images are some of them are quite dull and some of the images have distracting things.

In Figure 1 the clavicles are not attached to the shoulder joints. The clavicle can also be referred to as a collarbone which serves as a strut between the shoulder blade and the breastbone. The clips are used to attach them to prevent the shoulder dislocation. The 4$^{th}$ no rib of the patient is lying over the 5$^{th}$ rib it may be due to pregnancy. So, we cannot remove the clavicles part because if we remove that part it can remove prime features which can affect the ability of model to identify whether the X-ray belongs to a human which have normal chest conditions or not.

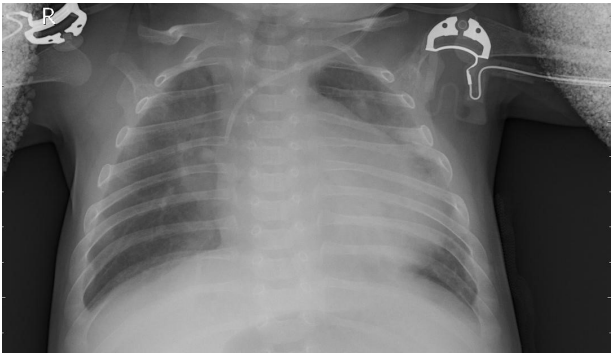
fig 1: Patient having clips in shoulder joint

In Figure 2 we have a drainage tube shown in our X-ray. The drainage pipe is used to treat the Pneumothorax. In Pneumothorax, water is stuck inside our lungs and drainage pipe is used to suck the water out from the lungs.

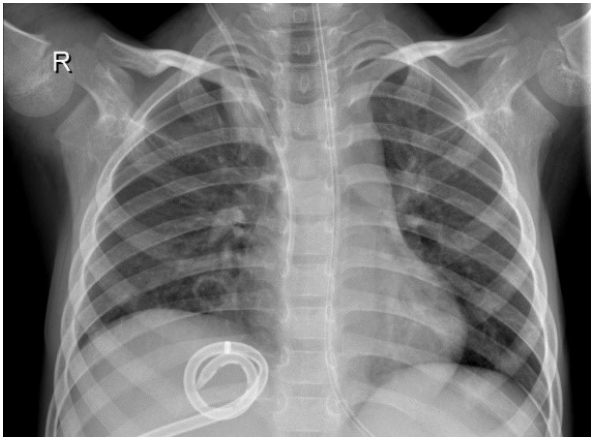
fig 2: Drainage Pipe inside x-ray

We cannot further pre-process it because it can remove the important factors through which disease can be identified from an x-ray, so we have to use it. There are many examples like this many patients have discs on their shoulder area. Some patients have Pacemakers on their hearts etc.

For preprocessing we use cropping because it is the only technique suitable for our task. After this, we resize images to 90 X 90 and convert the dimensions of the image to 2D Grayscale. We used both grayscale and RGB images to train our algorithm both have different results.

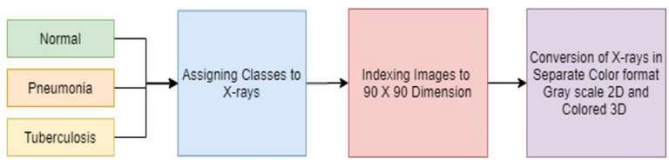
fig 3: Flow of preprocessing of the Dataset

### C. Label Assigning

After having all the X-ray images, we need to label them for our classifier to understand which Image belongs to the defined class, so we have three classes Normal, Pneumonia, and Tuberculosis and we label them as below in table 2.

Table 2: Label Distribution

| X-ray Type | Class |
|---|---|
| Normal | 0 |
| Pneumonia | 1 |
| Tuberculosis | 2 |

We used the python list in which we are adding an image with its Indexed folder using python function and then convert those lists into NumPy array.

### D. Generation of Pickle File

There are many approaches to do a certain task there are two approaches which can be used to do this task.

We can load Images in our IDE without generating a dataset pickle file, but it takes too much time and power by CPU and is not recommended.

We adopt this approach of Pickle file generation. So, pickle is a file which stores all the images of our dataset and we have separate Label file we use this approach because our system has not to use extra power to load all the images to the IDE in the pickle it is already done.

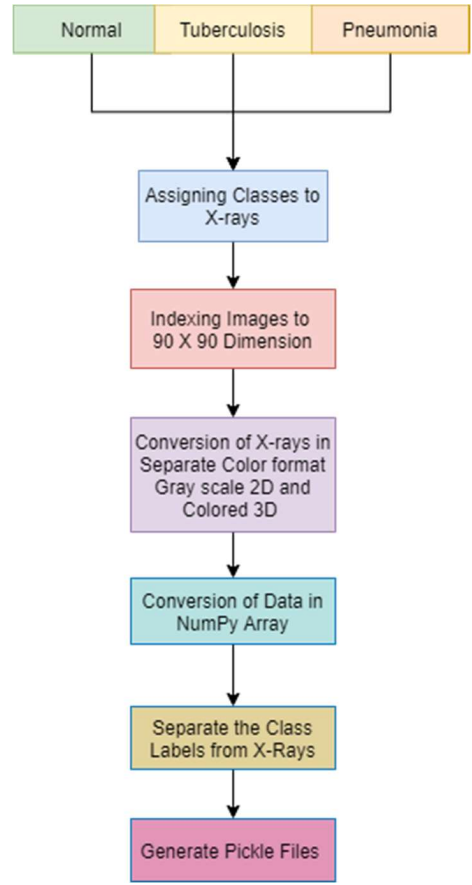
fig 4: Generation of Dataset

## IV. PROPOSED METHEDOLOGY

We are using a classification in our project. What our system is doing it is classifying the Xray from the three of the following categories.

- Normal
- Pneumonia
- Tuberculosis.

### A. Convolution Neural Network

Convolution Neural Networks best approach to Classify from Images. CNN is a type of Deep Neural networks mostly used to analyze visual imagery.

For example, CNN takes an image to classify it whether it is a dog or a cat. So, in case we have 2 classes Dog and cat with pictures representing each class we have to resize them all with fixed dimensions like (90 X 90 X 3) where 90 is representing Height and width and 3 is a dimension which means it is 3d image. IN CNN different layers are connected the input image is first of all pass through the Convolution layer. In figure 5 the architecture of our model is shown.

In the Conv layer, there is a set of filters applied on the image which decide the portion of the image to use then that portion with activation functions and then passed through the pooling layer. In the pooling layer, there are further filters to extract data. So, there is a choice to define the architecture by yourself by adding further layers and at the end the flatten layer is added which is passed through the last Conv layer having several classes with SoftMax activation function.

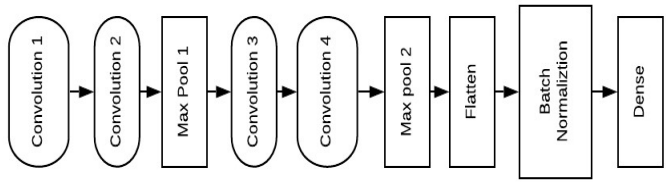

fig 5: Architecture of CNN Implemented

In fig 6 we have the visual representation of the equation of neural networks. The X1 represents the feature of the data. The data can be split into its features and its feature can be up to Xn each feature is multiplied with weights defined by neural network and then the result can is passed to the function which can produce the results. The function can be sigmoid, RelU, etc

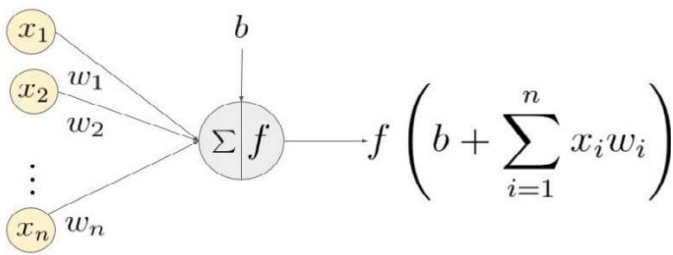

fig 6: Visual representation of equation of neural network[13]

### B. Vgg-16 based architecture of Neural Network

Vgg-16 is a Transfer learning algorithm. It is a convolution neural network which is differ in architecture having 16 convolution layers. The Vgg-16 is created in Image Net competition because of having its best results it got popular in the Deep learning. Now here is the difference. The CNN uses back propagation technique to maintain the weights to achieve good accuracy, but it uses predefined Image Net Weights. But in our case we are not using the Imagenet weights we are just using the VGG-16 architecture for CNN.

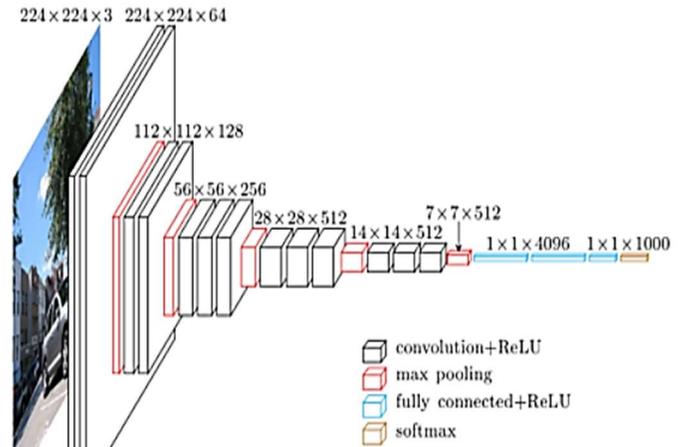

fig 6: Vgg-16 Architecture [11]

### C. Inception Network Based Architecture for CNN

Inception Layers or inception network is the state of art Deep learning architecture. They are used to allow researchers to perform efficient computation and deeper networks through a dimensionality reduction. It solves many problems like over fitting. The working of Inception layer is different it has 27 layers like a neural network but in this network the Convolution is performed on input using 3 types of convolution 1x1, 3x3, 5x5. Also, Max-pooling is performed and then outputs are concatenated and sent to next layer.

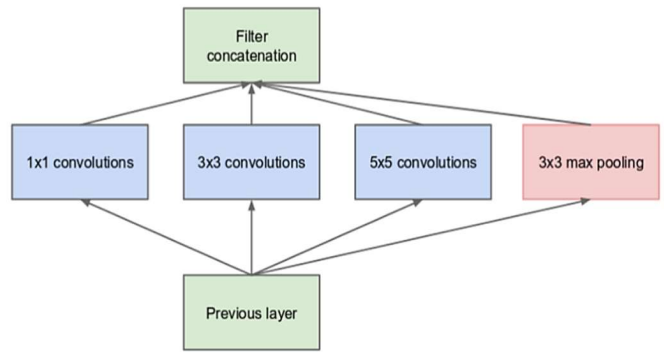

fig 7: Inception Net Architecture [12]

## V. RESULTS

### A. CNN Results

In Figure 5 we have presented the architecture of our primary algorithm with the best results achieved so far. This the architecture of Neural Network we used to achieve the best results on our data. Architecture is simple it consists of two convlolutional layers with max-pooling layers and again the convolution layer with max-pooling layers then we have a flatten layer to merge the data into one layer then we use batch normalization layer and at the end, we have a dense layer with softmax function.

In Figure 8 We have the training and validation graph of our neural network. This graph is showing us how the algorithm is fitting on the data accurately. The training accuracy is rising in every epoch but Validation accuracy results are not good as compare to the training ones this shows data is not overfitting because when training accuracy is lesser than validation it means our model is overfitting but in some epochs it is surpassing the training accuracy it doesn't mean it is overfitting it means the data in that epoch which our model is validated have easy examples in it that is why it is performing well on them. The data is trained in 10 epochs and the batch size is 120 so the starting accuracy of epoch is from 79% and then it improves from epoch to epoch.

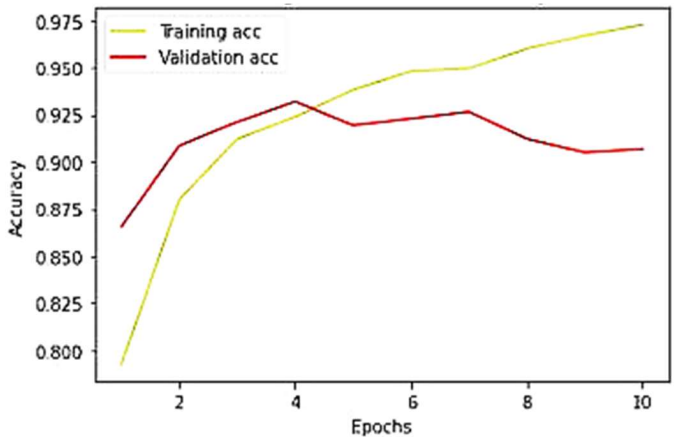

fig 8: Training and Validation Graph Accuracy

In Figure 9 we have training and validation loss graph of our neural network. This graph is showing us the loss in each epoch our model is having, and it shows our model is having loss decreased in every epoch and results are getting better and better. The loss of the training and validation is decreasing from epoch to epoch. As defined above the model is trained with 10 epochs having 0.2 validation split the model starting of loss is much higher than the last epoch.

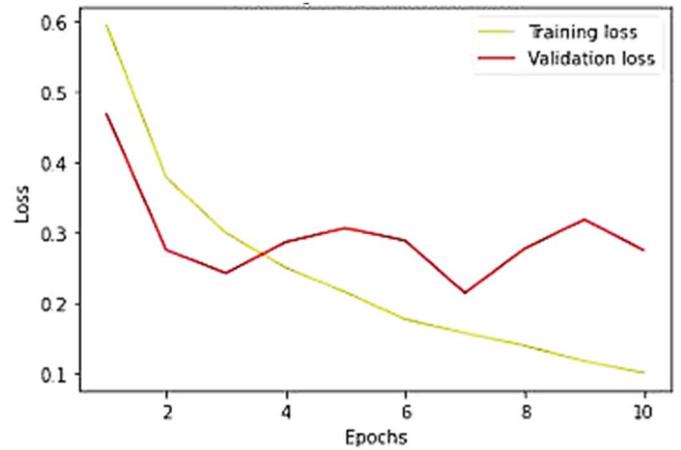

fig 9: Training and Validation Loss Graph

### B. Evaluation Metric

We used the Keras Evaluation function to evaluate our model the purpose of evaluation of our model is to get an overview how good our model is performing on the data which we provided it to get trained. So, the metric we used to evaluate our model is keras evaluation which give us the accuracy of 92.97% which means our model is performing good on the data.

```
1394/1394 [==============================] - 1s 392us/step
Evaluation 1
accuracy: 92.97%
```

fig 10: **Evaluation Accuracy**

### C. Deployment of model

We also deployed our machine learning model in production. To deploy it in production we choose web platform to deploy it. Further we used MVC web framework of python flask. The model is tested and evaluated using keras and then stored in a keras save function. We used the saved file and load it into flask project then pass the image to model by processing the image into the format which was given at the time of training of our model.